\documentclass[floatfix,aps,prb,preprint,showpacs,preprintnumbers,amsmath,amssymb,superscriptaddress]{revtex4-1}
\usepackage{epsfig}
\usepackage{graphicx}% Include figure files
\usepackage{color}
\usepackage{ulem}
\usepackage{graphicx,hyperref}
\usepackage{color,soul}

\begin{document}

\title{Nearly-free electrons in a $5d$ delafossite oxide metal}

\author{Pallavi Kushwaha}
\affiliation {Max Planck Institute for Chemical Physics of Solids, N{\"o}thnitzer Stra{\ss}e 40, 01217 Dresden, Germany}

\author{Veronika Sunko}
\affiliation {Max Planck Institute for Chemical Physics of Solids, N{\"o}thnitzer Stra{\ss}e 40, 01217 Dresden, Germany}
\affiliation {Scottish Universities Physics Alliance, School of Physics and Astronomy, University of St. Andrews, St. Andrews KY16 9SS, United Kingdom}

\author{P.~J.~W.~Moll}
\affiliation {Laboratory for Solid State Physics, ETH Zurich, Switzerland}

\author{L.~Bawden}
\affiliation {Scottish Universities Physics Alliance, School of Physics and Astronomy, University of St. Andrews, St. Andrews KY16 9SS, United Kingdom}

\author{J.~M.~Riley}
\affiliation {Scottish Universities Physics Alliance, School of Physics and Astronomy, University of St. Andrews, St. Andrews KY16 9SS, United Kingdom}
\affiliation{Diamond Light Source, Harwell Campus, Didcot, OX11 0DE, United Kingdom}

\author{Nabhanila Nandi}
\author{H.~Rosner}
\author{M.~P.~Schmidt}
\author{F.~Arnold}
\author{E.~Hassinger}
\affiliation {Max Planck Institute for Chemical Physics of Solids, N{\"o}thnitzer Stra{\ss}e 40, 01217 Dresden, Germany}

\author{T.~K.~Kim}
\author{M.~Hoesch}
\affiliation{Diamond Light Source, Harwell Campus, Didcot, OX11 0DE, United Kingdom}

\author{A.~P.~Mackenzie}
\email{Andy.Mackenzie@cpfs.mpg.de}
\affiliation {Max Planck Institute for Chemical Physics of Solids, N{\"o}thnitzer Stra{\ss}e 40, 01217 Dresden, Germany}
\affiliation {Scottish Universities Physics Alliance, School of Physics and Astronomy, University of St. Andrews, St. Andrews KY16 9SS, United Kingdom}

\author{P.~D.~C.~King}
\email{philip.king@st-andrews.ac.uk}
\affiliation {Scottish Universities Physics Alliance, School of Physics and Astronomy, University of St. Andrews, St. Andrews KY16 9SS, United Kingdom}

             \begin{abstract}
Understanding the role of electron correlations in strong spin-orbit transition-metal oxides is key to the realisation of numerous exotic phases including spin-orbit assisted Mott insulators, correlated topological solids, and prospective new high-temperature superconductors. To date, most attention has been focussed on the $5d$ iridium-based oxides. Here, we instead consider the Pt-based delafossite oxide PtCoO$_2$. Our transport measurements, performed on single-crystal samples etched to well-defined geometries using focussed ion-beam techniques, yield a room-temperature resistivity of only 2.1~$\mu\Omega$cm, establishing PtCoO$_2$ as the most conductive oxide known. From angle-resolved photoemission and density-functional theory, we show that the underlying Fermi surface is a single cylinder of nearly hexagonal cross-section, with very weak dispersion along k$_z$. Despite being predominantly composed of $d$-orbital character, the conduction band is remarkably steep, with an average effective mass of only 1.14$m_e$. Moreover, the sharp spectral features observed in photoemission remain well-defined with little additional broadening for over 500~meV below E$_F$, pointing to suppressed electron-electron scattering. Together, our findings establish PtCoO$_2$ as a model nearly-free electron system in a $5d$ delafossite transition-metal oxide.
\end{abstract}

\date{\today}% It is always \today, today,
            
\maketitle 

\section{Introduction}
The delafossite structural series of oxides has recently attracted considerable attention because of the remarkable and varied properties of the compounds in the series.  Its general formula is ABO$_2$, in which A is a noble metal (Pt, Pd, Ag or Cu) and B a transition metal (e.g. Cr, Co, Fe, Al and Ni)~[\onlinecite{Shannon, Rogers, Prewitt, Marquardt}].  The metal atoms are found in layers with triangular lattices, stacked along the perpendicular direction in various sequences, leading to structures similar to the layered rocksalt structure of the well-known ionic conductor LiCoO$_2$ and of NaCoO$_2$, which superconducts when intercalated by water.  Interlayer coupling is weak, so the delafossites are quasi-two dimensional.  

The low temperature properties across the series vary considerably with A and B combinations.  Known materials include candidate magnetoelectric insulators and thermoelectrics~[\onlinecite{Singh}, \onlinecite{Wang}], transparent conductors~[\onlinecite{Yanagi}] and band insulators~[\onlinecite{Shannon, Rogers, Prewitt, Marquardt}].  There are also intriguing metals such as AgNiO$_2$ and PdCrO$_2$ in which the conduction takes place between Mott insulating layers with magnetic order~[\onlinecite{Takatsu 2009, Takatsu 2010, Ok, Coldea, Takatsu 2014, Hicks 2015}], and non-magnetic metals with ultra-high conductivity, such as PdCoO$_2$.  Although they have been less studied than the layered perovskites, the delafossites have considerable potential for new physics and technology; epitaxial multilayers of delafossites and of delafossite/rocksalt combinations can be imagined. There is, therefore, a strong motivation to understand the fascinating properties of individual materials in the series.  

In PdCoO$_2$, cobalt has formal valence of 3+ and the $3d^6$ configuration, meaning that the states at the Fermi level have dominantly Pd character~[\onlinecite{Eyert, Seshadri, Kim, Noh_orb}], with a single half-filled band crossing the Fermi level.  At 300~K the resistivity ($\rho$) is 2.6~$\mu\Omega$cm, lower per carrier than that of Cu~[\onlinecite{Hicks}].  At low temperatures both the shape and the value of $\rho$ are very unusual.  It is best fitted by $\rho=\rho_0+\beta{e^{-T^*/T}}$ (with $\beta$ a constant and $T^*=165$~K) rather than the power law expected for standard phonon scattering.  At least as surprising as the form of $\rho$ is its value.  A residual resistivity of only 7.5~n$\Omega$cm has been reported~[\onlinecite{Hicks}].  This corresponds to a mean free path ($\ell$) of 20~$\mu$m, or $\sim\!10^5$ lattice spacings, an astonishing value in crystals grown from fluxes in hot crucibles that have not been subject to any post-growth purification.  The high value of $\ell$ in PdCoO$_2$ has important consequences for a number of physical properties, for example the out of plane magnetoresistance which is huge and varies strongly with field angle~[\onlinecite{Takatsu 2013}], behavior consistent with observation of the long sought `axial anomaly'~[\onlinecite{Kikugawa 2014}, \onlinecite{Goswami 2015}]. 

The subject of this paper, PtCoO$_2$, offers an opportunity to extend the study of the intriguing metallic delafossite oxides to a $5d$ system where spin-orbit coupling (SOC) is considerably stronger. Although crystal growth~[\onlinecite{Prewitt}] and room temperature resistivity~[\onlinecite{Rogers}] were reported over forty years ago, no temperature-dependent transport studies were carried out on those crystals, and further crystal growth proved to be extremely challenging.   Here we describe crystal growth using a different method to those previously attempted in PtCoO$_2$. This yields high-quality single crystals large enough for spectroscopic measurements. From angle-resolved photoemission (ARPES), combined with density-functional theory calculations and precise transport measurements performed on focused ion beam patterned crystals, we show that PtCoO$_2$ hosts ultra-high conductivity derived from a single half-filled and nearly-free-electron-like Pt $5d$ band.

\section{Results}
\subsection{Single crystals}
First, we describe our crystal growth.  Shannon \textit{et. al}~[\onlinecite{uspatent}] reported single crystal growth of Pt$_x$Co$_y$O$_2$ (where x and y were $0.85~\pm~0.15$) several decades ago.  Stoichiometric PtCoO$_2$ was grown only under high pressure (3000 atm), with other growth conditions resulting only in nonstoichiometric crystals. Tanaka \textit{et. al}~[\onlinecite{Tanaka}] were able to grow stoichiometric PtCoO$_2$ by using a metathetical reaction under vacuum, but the crystal size was limited to $30~\mu$m. Here we used a technique similar to that applied in Ref.~[\onlinecite{Takatsu 2007}] to the growth of PdCoO$_2$ (see methods) to realise large stoichiometric PtCoO$_2$ single crystals.

%FIGURE 1
\begin{figure}[t]
	\begin{center}
	\includegraphics[width=12 cm]{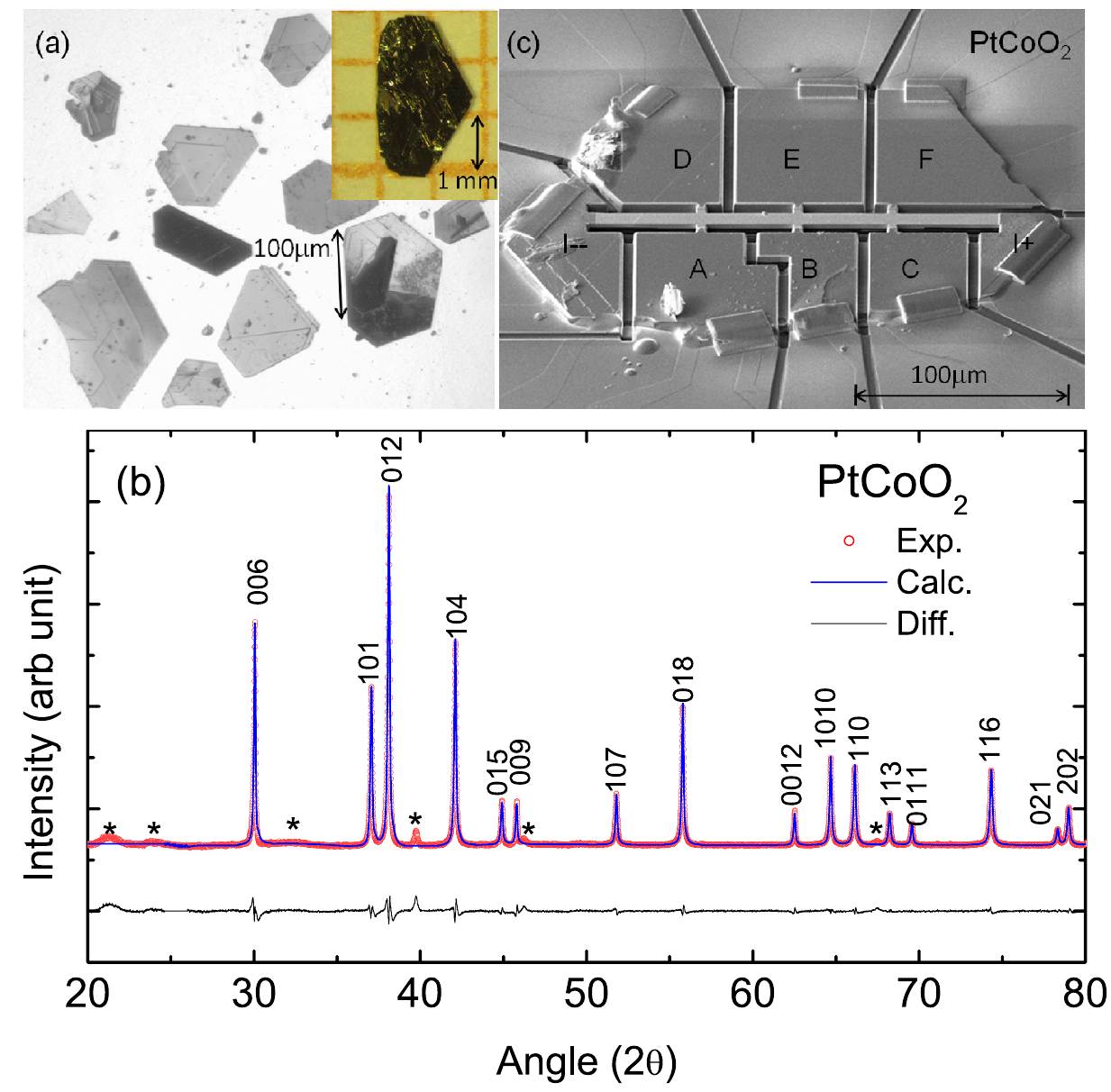}
	\end{center}
	\caption{ \label{f1} {\bf Single-crystal PtCoO$_2$ samples.} (a) Optical microscope image of as-grown crystals of PtCoO$_2$. (b) Le-Bail fitting of powder x-ray diffraction pattern along with fitted curve, and the difference curve. All peaks are labeled with corresponding $hkl$ values. Peaks marked with * correspond to unavoidable unreacted PtCl$_2$ stuck to the crystal surface. (c) SEM image of a sample used for transport measurements in which a focused ion beam was used to define a measurement track of well-defined geometry.}
\end{figure}

Fig.~\ref{f1}(a) shows optical pictures of the as-grown crystals. They form as triangular or hexagonal plates. Terrace-type lateral growth leads to variation in crystal thickness from one side to another as evident from Fig.~\ref{f1}(a). However there were many crystals with uniform thickness without any steps. Due to their brittleness and layered nature, the typical size of these crystals varies from in-plane dimensions of a few $\mu$m to 0.3~mm, with a maximum thickness of $3~\mu$m. Bigger crystals were grown after employing modified techniques with more complex temperature profiles~[\onlinecite{Pallavi2015}]. A cluster of such big as-grown crystals is shown in the inset of Fig.~\ref{f1}(a). These crystals were used for the ARPES measurements discussed below. Single crystals were separated mechanically using a scalpel, and then separated from unreacted CoO and from CoCl$_2$ powder by cleaning the product with boiling alcohol. All crystals were characterized by a scanning electron microscope (SEM) with an electron probe micro analyzer (EPMA) and confirmed to be single phase and have no inclusions. Le-Bail fitting of the powder x-ray diffraction pattern (XRD) was performed using space group $R$$\bar{3}$$m$ (space group no.~166). The experimental XRD pattern along with the fitted pattern and the difference between the two are shown in Fig. 1~(b). The refined lattice parameters are $a=2.82259(\pm5)$~\AA\ and $c=17.8084(\pm3)$~\AA.

\subsection{Transport measurements}
%FIGURE 2
\begin{figure}[b]
	\begin{center}
	\includegraphics[width=10 cm]{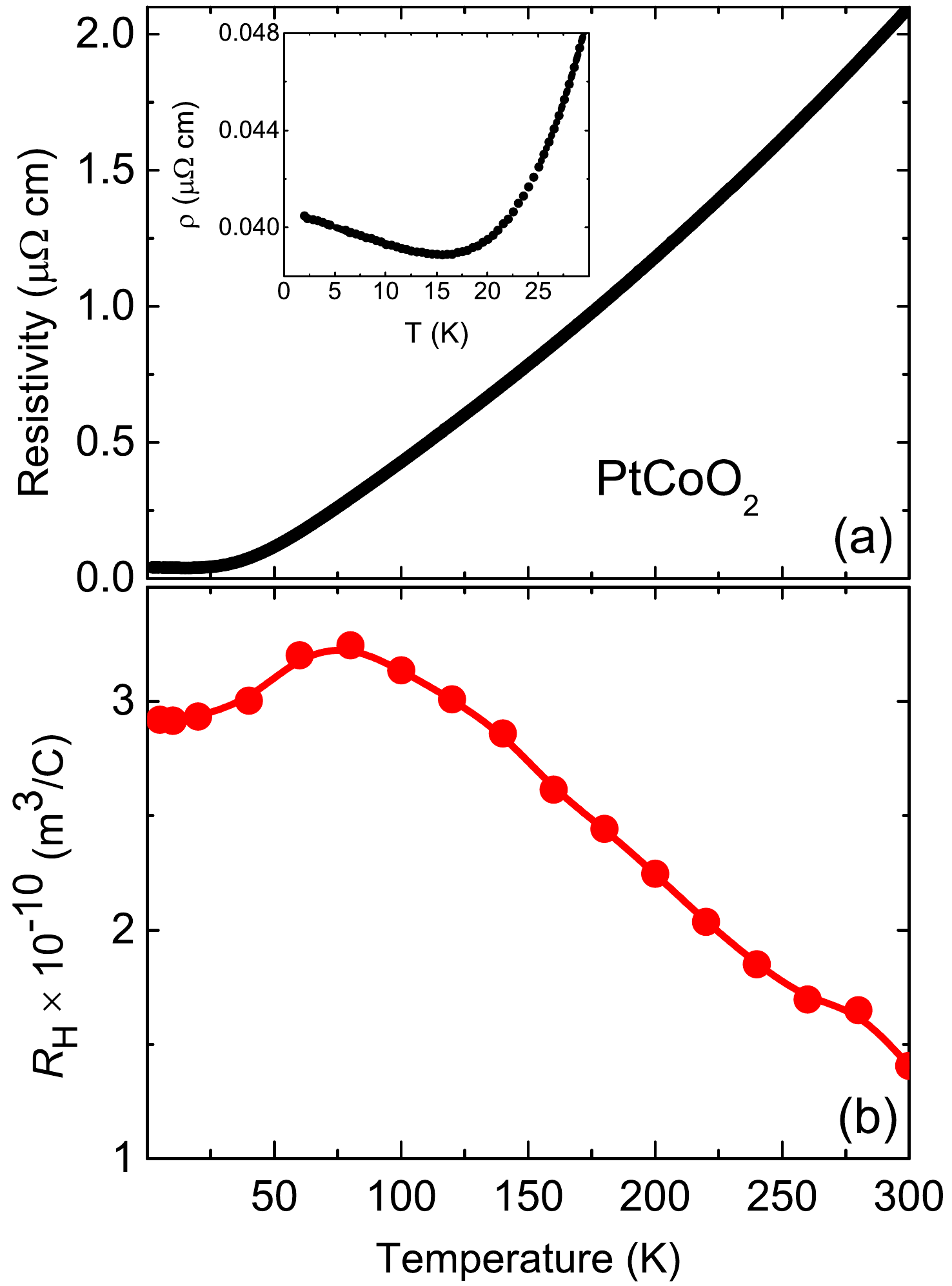}
	\end{center}
	\caption{ \label{f2}  {\bf Temperature-dependent transport.} (a) The temperature dependent in-plane resistivity of PtCoO$_2$ in zero applied magnetic field. {\color{black}The inset shows a magnified view of the low-temperature resistivity revealing an upturn below 16~K. (b) The temperature dependence of the Hall coefficient ($R_H$), calculated by taking the field gradient between 7 and 9~T of the data shown in Supplementary Fig.~S1.}}
\end{figure}

Since the resistivity of PtCoO$_2$ is very low~[\onlinecite{Rogers}], it was challenging to obtain accurate absolute values of resistivity due to uncertain geometrical factors.  To overcome this, and to enhance the precision with which we could measure even smaller resistances at low temperatures, we made use of focused ion beam techniques to prepare samples with well-defined geometries~[\onlinecite{Moll}].  One example is shown, along with its dimensions (length between two voltage contacts 40~$\mu$m, width 8.4~$\mu$m and thickness 2.6~$\mu$m), in Fig. 1(c).  

In Fig.~\ref{f2} we show the in-plane resistivity of PtCoO$_2$ crystals from the growth run described above.  In total, approximately ten crystals were studied, and the reproducibility of the temperature dependence was excellent.  The data are from the well-defined microstructure shown in Fig.~\ref{f1}, allowing the absolute value of resistivity to be determined with an accuracy of better than 5\%.  {\color{black}At room temperature, $\rho=2.1$~$\mu\Omega$cm, the lowest ever measured in an oxide metal. This is 20\% lower than that of PdCoO$_2$~[\onlinecite{Hicks}] and approximately a factor of 4~[\onlinecite{Pearsall}] and 12~[\onlinecite{Ryden}] lower than in ReO$_3$ and IrO$_2$, respectively, which are themselves famous examples of good $5d$ oxide conductors.}  The resistivity decreases to less than 40~n$\Omega$cm at 16~K, but then rises again by approximately 4\% between 16~K and 2~K  {\color{black}as shown in the inset of Fig.~\ref{f2} (a)}. The residual resistivity ratio $\rho_{300K}/\rho_{15K}$ is typically 50-60.  The upturn is observed consistently in both as-grown crystals and those that were microstructured with the focused ion beam, so we are certain that it is not the result of ion-beam induced disorder.  The measured magnetoresistance of PtCoO$_2$ is positive for all temperatures above 2~K, making it unlikely that the Kondo effect is the cause of the upturn. Its origin will be the subject of further investigation.

 {\color{black}In Fig.~\ref{f2} (b)} we show the measured Hall coefficient deduced from measurements of the transverse resistivity of the device shown in Fig.~\ref{f1}(c) {\color{black}(see also Supplementary Fig.~S1)}. Below 20~K, the data are dominated by the high-field regime in which the Fermi surface volume alone is expected to determine the value of the Hall coefficient~[\onlinecite{Hurd}]. The value in that region is approximately $2.9\times10^{-10}$~m$^3$/C, close to the value that we measure for PdCoO$_2$ under the same conditions (data not shown), and within 10\% of the expectation for a single band containing one electron per Pt.

\subsection{Electronic structure}
%FIGURE 3
\begin{figure}[t]
	\begin{center}
	\includegraphics[width=14 cm]{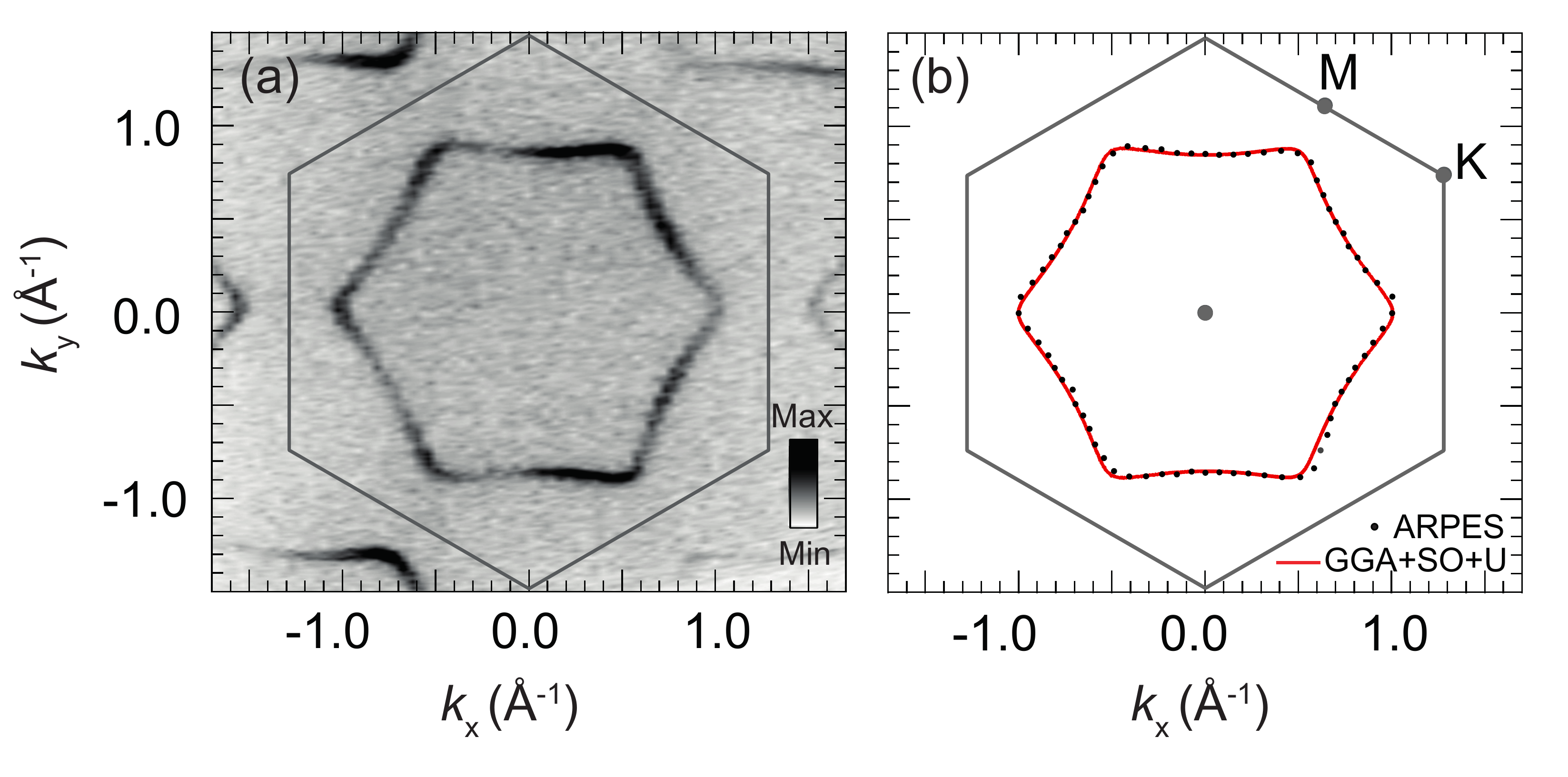}
	\end{center}
	\caption{ \label{f3} {\bf Single-band faceted Fermi surface.} (a) Fermi surface of PtCoO$_2$ measured by ARPES integrated over $E_F\pm{5}$~meV. The solid line represents the Brillouin zone. The Fermi surface area is 8\% smaller than would be expected for a half-filled band, as discussed in the main text, but its shape is in excellent agreement with the Fermi surface obtained from {\color{black}GGA+SO+U} calculations ($U=4$~eV), shown in red in (b) and scaled to match the experimental area. The dots represent the Fermi momenta extracted from (a) by fitting MDCs radially around the measured Fermi surface.}
\end{figure}
Our ARPES measurements (Fig.~\ref{f3}) directly reveal just such a Fermi surface, comprised of a single electron pocket centred at the $\Gamma$ point of the Brillouin zone. The sharp spectral features are indicative of quasi-two-dimensional electronic states~[\onlinecite{Riley}]. The near-hexagonal Fermi surface, rotated by 30$^\circ$ with respect to the Brillouin zone, is qualitatively similar to that of the bulk Fermi surface of PdCoO$_2$~[\onlinecite{Noh}]. We find, however, that the PtCoO$_2$ Fermi surface has sharper corners and slightly concave faces. This is consistent with previous first-principles calculations~[\onlinecite{Ong and Singh}], as well as our own density-functional theory results (Fig.~\ref{f3}(b)). {\color{black}Our calculations predict an additional hole band in the vicinity of $E_F$ at the zone edge (Fig.~\ref{figDOS}(a)), whose position depends delicately on the oxygen position (0,0,$z$) as shown in Supplementary Fig.~S2(b). Neglecting SOC, this intersects the Fermi level when using either the experimental~[\onlinecite{Opos}] or fully-relaxed oxygen positions, creating additional small lens-shaped Fermi surface pockets. Including SOC, however, lowers these bands by $\sim\!100$~meV, pushing them below the Fermi level. Spin-orbit interactions therefore stabilise a robust single-band Fermi surface in PtCoO$_2$, consistent with our ARPES measurements as well as recent de-Haas van Alphen results~[\onlinecite{dHvA}]. }

%FIGURE 4
\begin{figure}[t]
        \begin{center}
        \includegraphics[width=10 cm]{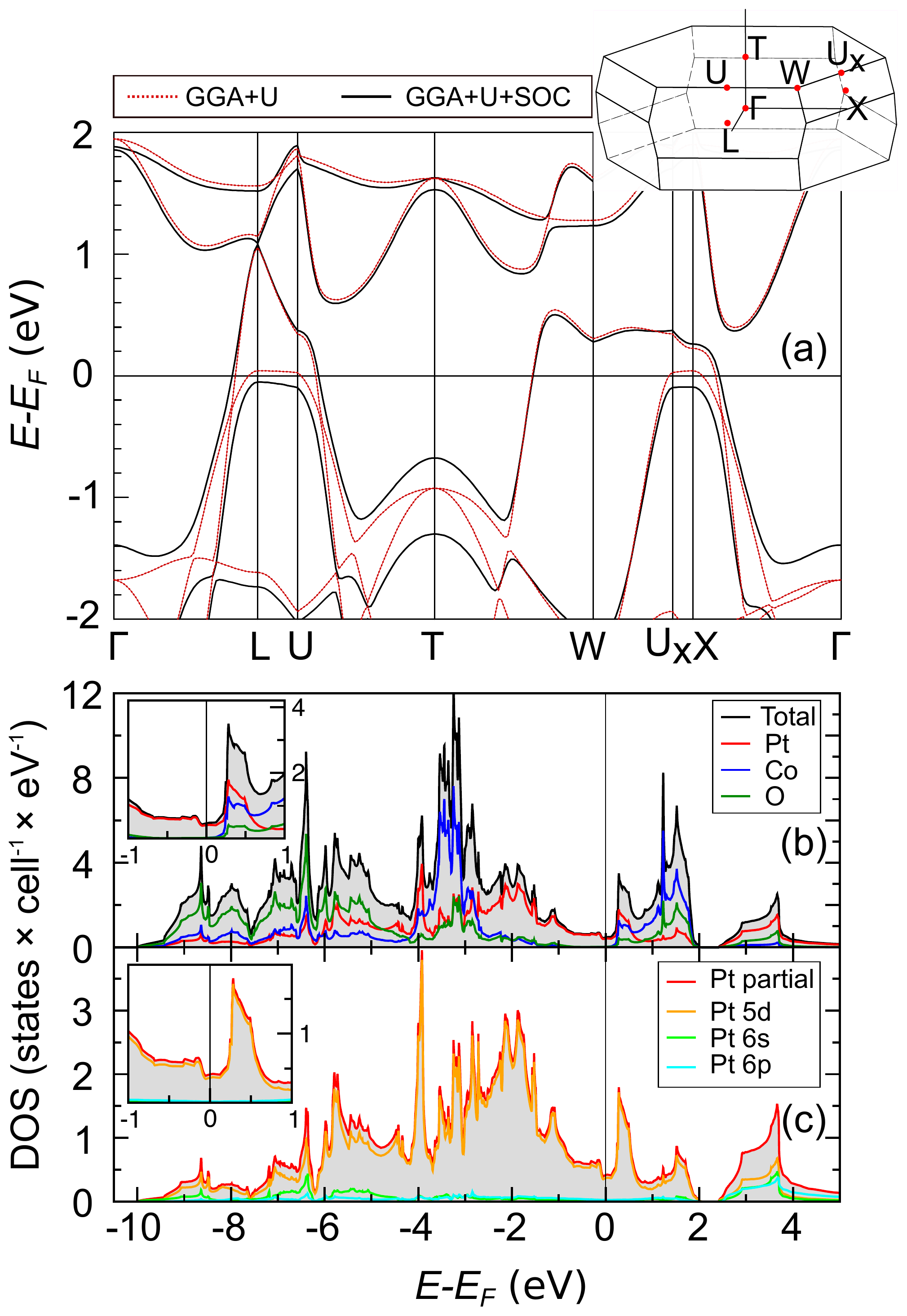}
        \end{center}
        \caption{ \label{figDOS} {{\color{black}{\bf Electronic structure.} (a) Calculated bulk electronic structure (see methods) with and without spin-orbit coupling. The high-symmetry points are labelled on the bulk Brillouin zone shown inset. Inclusion of SOC pushes the hole bands at, e.g., the L-point below $E_F$, while leaving the remaining bands crossing the Fermi level almost unchanged. (b) Partial and (c) orbitally-resolved density of states (including SOC), revealing the states at E$_F$ to originate almost exclusively from Pt-5d.}}
}
\end{figure}
%FIGURE 5
\begin{figure}[b]
	\begin{center}
	\includegraphics[width=12 cm]{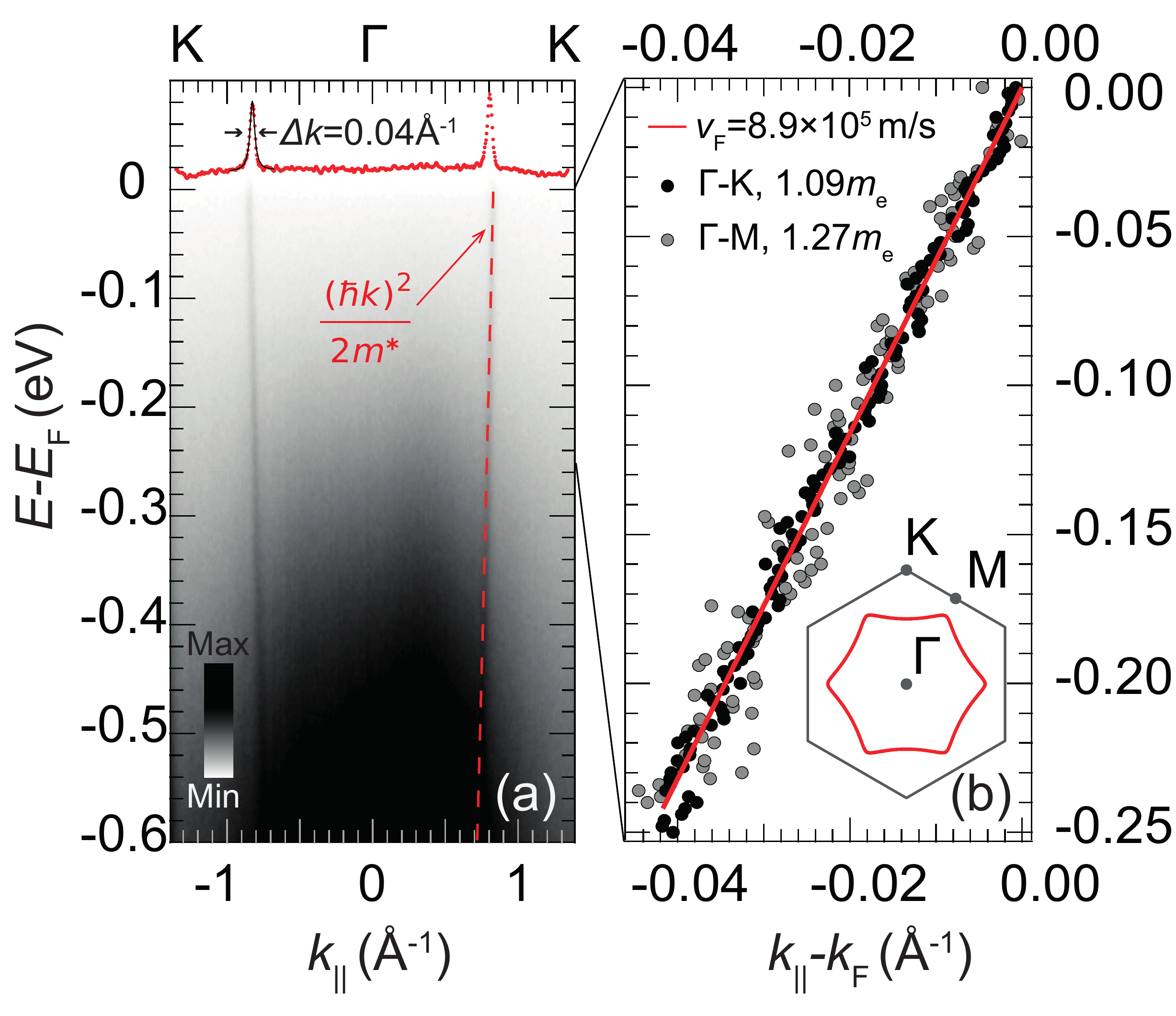}
	\end{center}
	\caption{ \label{f5} {\bf Weakly-interacting quasiparticle dispersion.} (a) Electronic structure along the $\Gamma-K$ direction. The single band crossing the Fermi level can be traced down to more 0.5~eV binding energy with little broadening. The dashed red line corresponds to a parabolic dispersion with an effective mass  $m^*=1.09m_e$. An MDC at the Fermi level ($E_F\pm6$~meV) is shown by the dots, with a fit to a Lorentzian peak indicating a FWHM of $0.04$~\AA$^{-1}$. (b) The grey and black dots represent the peak positions of the fits to the MDCs  along the $\Gamma-M$ and $\Gamma-K$ direction, respectively. A linear fit to each data set independently yields a Fermi velocity of  $8.9\times10^5$~m/s (shown by the solid line), giving an effective mass of $1.09m_e$ along $\Gamma-K$ and $1.27m_e$ along $\Gamma-M$.}
\end{figure}

The Fermi surface area we extract from ARPES, by fitting momentum distribution curves (MDCs) as a function of angle around the Brillouin zone, is $2.6\pm0.1$~\AA$^{-2}$. The corresponding Luttinger count, assuming perfect two-dimensionality, is $n=(0.92\pm0.05)e^-$. This is comparable to ARPES measurements of PdCrO$_2$~[\onlinecite{Sobota}], but is slightly smaller than that expected for a half-filled zone derived from a single conduction electron per Pt. In contrast, de Haas-van Alphen measurements indicate a Luttinger count within {2\%} of half-filling. We thus ascribe the deviations in our ARPES measurements as due to a possible small difference in surface chemistry from the bulk, sample-to-sample stoichiometry variations, or surface charge transfer. Nonetheless, we find almost perfect agreement of the shape of our measured Fermi surface with our calculations from density-functional theory scaled to have the same area (Fig.~\ref{f3}(b)). We therefore conclude that our measurements are indicative of the bulk electronic structure of PtCoO$_2$. {\color{black}We note that, at lower photon energies, weak spectral features emerge giving rise to shoulders on the states reported here and residual spectral weight at the zone centre. From comparison to previous surface electronic structure calculations for PdCoO$_2$~[\onlinecite{Kim}] and ARPES measurements of PdCrO$_2$~[\onlinecite{Sobota}], we attribute these to Pt-terminated surface states not visible at the higher photon energies shown here due to photoemission cross-section variations~[\onlinecite{Noh2}], although their weak weight precludes an unambiguous assignment between these features. Other cleaves showed stronger surface-derived features but a bulk band in quantitative agreement with the one found here.} 

{\color{black}Fig.~\ref{f5}(a) shows the bulk band dispersion} measured along the $\Gamma-K$ direction. Consistent with our measured Fermi surface, we find a single band in the vicinity of $E_F$. It is extremely steep, dispersing by less than 5\% of the Brillouin zone over the energy range of 0.6~eV below the Fermi level shown in Fig~\ref{f5}(a). Indeed, from fits to MDCs, we find negligible deviations from linearity over a range exceeding 0.25~eV below $E_F$, together with a high Fermi velocity of $(8.9\pm0.5)\times10^5$~m/s [$5.8\pm0.3$~eV\AA/$\hbar$]. This is several times larger than observed in other metallic $5d$ oxides~[\onlinecite{dlT}, \onlinecite{Nie}], and yields an effective mass, $m^*=1.09m_e$, within 10\% of the free-electron mass. With no free parameters, employing this effective mass within a simple parabolic model well reproduces our measured dispersion over an extended energy range (Fig.~\ref{f5}(a)). 

The Fermi velocity [$(8.9\pm0.9)\times10^5$~m/s] that we extract along $\Gamma-M$ is the same as that along $\Gamma-K$ within our experimental resolution (Fig.~\ref{f5}(b)). This is almost 20\% higher than that measured along the $\Gamma-M$ direction in PdCoO$_2$~[\onlinecite{Noh}]. Due to the higher $k_F$ along $\Gamma-M$ than $\Gamma-K$, the effective mass obtained from our measured Fermi velocity, $m^*=1.27m_e$, deviates further from the free-electron mass. The average mass of $m^*=1.18m_e$ is in excellent agreement with the value of $m^*=(1.14\pm0.05)m_e$ obtained from quantum oscillations of PtCoO$_2$~[\onlinecite{dHvA}], and again notably smaller than the values of around $1.5m_e$ obtained for PdCoO$_2$~[\onlinecite{Hicks}]. 

\section{Discussion}
The above measurements point to light quasiparticles in this system, in keeping with their extremely high conductivity, with an even wider electronic bandwidth of the state crossing $E_F$ in PtCoO$_2$ than the Pd-derived sister compound. In PdCoO$_2$, it has proved controversial whether the high conductivity and low effective masses are mediated by significant Pd $s$--$d$ orbital mixing~[\onlinecite{Rogers}, \onlinecite{Eyert}, \onlinecite{Hicks}, \onlinecite{Noh}, \onlinecite{TanakaPB}].
To specifically address this issue, we have performed our density functional calculations using a local-orbital rather than plane-wave basis (see methods), which allows a well-justified assignment of orbital characters. We find that the density of states at the Fermi level originates predominantly from Pt states, with almost 90\% Pt-related character.  Approximately 85\% of these states belong to the Pt~5$d$ orbitals with only a small admixture of Pt~$6s$ and Pt~6$p$ (Fig.~\ref{figDOS}(b)). Due to the localized nature of the Co $3d$ shell, the strong Coulomb repulsion for these states was taken into account explicitly (see Fig.~\ref{figDOS} (a) and methods). This suppresses an unphysical hybridization of the Co-$3d$ states with the Pt-$5d$ states near the Fermi level which cause a pronounced rounding of alternating Fermi surface corners in contrast to the experiment. This suppression occurs for $U_{3d}\gtrsim3$~eV, after which the Fermi surface topology and its orbital character are insensitive to the value of $U_{3d}$ used over our investigated range of up to 6~eV. The Pt $d$-orbital derived nature of the Fermi surface found here is entirely in keeping with its faceted nature that we observe, while the increase of electronic bandwidth compared to PdCoO$_2$ can naturally be understood from the spatially more extended nature of Pt~$5d$ than Pd~$4d$ orbitals. Remarkably, the effective masses we observe here are a factor of 2-3 smaller than even those of the strongly \textit{s}-\textit{d}-hybridized electron Fermi surface sheets in Pt metal~[\onlinecite{Ketterson}]. 

From the bulk Fermi surface, we can straightforwardly convert our resistivity measurements shown in Fig.~\ref{f2} to a temperature-dependent mean free path, which we find rises from 700~\AA{} at room temperature to as high as 4~$\mu$m at 16~K. This is qualitatively consistent with the sharp spectral features we observe in ARPES. We observe narrow line widths of MDCs at the Fermi level of better than 0.04~\AA$^{-1}$ (Fig.~\ref{f5}(a)), very similar to those reported previously for PdCoO$_2$~[\onlinecite{Noh}] and comparable to our experimental angular resolution. These remain clearly resolved to high binding energies, with $\lesssim\!\!0.01$~\AA$^{-1}$ increase in linewidth over the first 350~meV binding energy range. This indicates a striking resilience of the quasiparticle lifetime away from the Fermi level, suggesting weak electron-electron scattering in PtCoO$_2$. This is in contrast to other well-known Fermi-liquid oxide metals such as Sr$_2$RuO$_4$~[\onlinecite{Damascelli}] and Sr$_2$RhO$_4$~[\onlinecite{Baumberger}], which host sharp spectral features at low energy, but rapidly develop substantial broadening away from $E_F$. Additionally, those systems host moderate correlation-induced enhancements of the effective mass. The lack of both of these features here points to a {\color{black}particularly} weak influence of electron-electron interactions on the electronic structure of PtCoO$_2$. 

This is not the situation in all {\color{black}quasi two-dimensional} $5d$ transition-metal oxides. In the layered iridates, for example, electronic correlations are still strong enough to drive a Mott-like transition to an insulating state once structural distortions and strong spin-orbit coupling collectively narrow the electronic bandwidth~[\onlinecite{Nie}, \onlinecite{BJKim}, \onlinecite{King}]. {\color{black}In contrast, itineracy is preserved in PtCoO$_2$ by the isolated single-band nature of the Fermi surface. In the absence of band-folding arising from structural distortions, not present in PtCoO$_2$, this band remains relatively unaffected by SOC, while large hybridization gaps are only opened away from $E_F$ where band crossings between different orbital character states occur (see, e.g., around the T point in Fig.~\ref{figDOS}(a) where a $d_{xz/yz}$ degeneracy is lifted by approximately 600~meV). Thus the bandwidth of the state crossing $E_F$ in PtCoO$_2$ remains large, away from the narrow-band $J_{eff}=1/2$ limit of the iridates. This naturally renders the electronic correlations expected of a $5d$ system insufficient to drive a Mott transition. Nonetheless, the extremely good metallicity and lack of spectral broadening away from $E_F$ point not only to large bare bandwidths, but also to suppressed electron-electron interactions compared to other wide-band transition-metal oxides.  Indeed, the fact that the Fermi velocities shown in Fig.~\ref{f5}(b) are so close to the free electron values is itself remarkable, because the highly faceted Fermi surface demonstrates the influence of the lattice potential on the states near $E_F$. Nonetheless, our measurements firmly establish PtCoO$_2$ as a model half-filled nearly-free-electron oxide metal, with high-quality crystals rendering this system an ideal test-bed for future studies of the nature of electronic correlations and the origins of extremely good metallicity in both Pt- and Pd-based delafossite oxides.}

\section*{Materials and Methods}
{\bf Single-crystal growth:} Powders of reagent grade PtCl$_2$ ($99.99+\%$ purity, Alfa Aesar) and CoO ($99.995\%$ purity, Alfa Aesar) were ground together for approximately an hour under an inert atmosphere in accordance with the chemical reaction PtCl$_2$ + 2CoO$\rightarrow$ PtCoO$_2$ + CoCl$_2$. The mixed powder was then sealed in a quartz tube under a vacuum of $5\times10^{-6}$ Torr. The sealed quartz tube was heated in a vertical furnace to 800~$^{\circ}$C for 5 hours and cooled down to 740~$^{\circ}$C at a rate of 7.5~$^{\circ}$C/hour and kept at this temperature for 30 hours. Finally, the furnace was cooled from 740~$^{\circ}$C to room temperature at a rate of 90~$^{\circ}$C/hour. 

{\bf Transport:} Transport measurements were performed using standard four-probe a.c. techniques in $^4$He cryostats (Quantum Design), with measurement frequencies in the range 50--200 Hz, magnetic fields of up to 14~T and the use of single-axis rotators.  The Hall effect was studied using reversed-field sweeps in the range $-9\;\mathrm{T}<B< 9\;\mathrm{T}$ at a series of fixed temperatures. 

{\bf Angle-resolved photoemission:} ARPES measurements were performed using the I05 beamline of Diamond Light Source, UK. Samples were cleaved {\it in-situ} at the measurement temperature of $\sim\!6$~K. Measurements shown in Fig.~\ref{f3} and Fig.~\ref{f5} were performed using 110~eV and 118~eV $p$-polarised light, respectively.  A Scienta R4000 hemispherical analyser was used for all the measurements.

{\bf Density-functional theory calculations:} Relativistic density 
functional (DFT) electronic structure calculations were performed 
using the full-potential FPLO code [\onlinecite{fplo}], version 
fplo14.00-47. For the exchange-correlation potential, within the 
general gradient approximation (GGA), the parametrization of 
Perdew-Burke-Ernzerhof [\onlinecite{PBE}] was chosen.  {\color{black}The spin-orbit coupling was treated non-perturbatively solving the four component Kohn-Sham-Dirac equation [\onlinecite{KSD}].} To obtain 
precise band structure and Fermi surface information, the  
calculations were carried out on a well converged mesh of  27000  
$k$-points (30x30x30 mesh, 2496 points in the irreducible wedge of  
the Brillouin zone). The strong Coulomb repulsion in the Co-3d shell 
was taken into account in a mean field way applying the GGA+$U$ 
approximation [\onlinecite{LDU}] in the atomic-limit-flavor (AL). For 
all calculations, the experimental lattice parameters have been used.

\section*{Acknowledgements}

We thank SCOPE-M and Philippe Gasser at ETH Zurich for supporting the FIB work. This work was supported by the Engineering and Physical Sciences Research Council, UK (grant number EP/I031014/1). PDCK acknowledges support from the Royal Society through a University Research Fellowship. We thank Diamond Light Source for access to beamline I05 that contributed to the results presented here.\\
\
\\
Competing Interests: The authors declare that they have no competing interests.\\
\
\\
Data underpinning this publication can be accessed at http://dx.doi.org/10.17630/7a88d794-d747-4aa2-a099-9335c5b702ec \\
\
\\
Author Contributions: PK and MPS performed the crystal growth; VS, PK, PJWM, LB, JMR, NN, FA, EH, and PDCK performed the experimental measurements; HR performed the theoretical calculations; TKK and MH built and maintained the ARPES end station and provided experimental support; APM and PDCK were responsible for overall project planning and direction and wrote the paper with input and discussion from all co-authors.

\section*{Supplemental Materials}
\noindent Fig.~S1: Field-dependent Hall effect measurements\\
\noindent Fig.~S2: Additional density-of-states calculations

%\bibliographystyle{phaip}
% \bibliography{References}

\begin{thebibliography}{10}
\expandafter\ifx\csname url\endcsname\relax
  \def\url#1{\texttt{#1}}\fi
\expandafter\ifx\csname urlprefix\endcsname\relax\def\urlprefix{URL }\fi
\providecommand{\bibinfo}[2]{#2}
\providecommand{\eprint}[2][]{\url{#2}}

\bibitem{Shannon} R. D. Shannon, D. B. Rogers, and C. T. Prewitt,Chemistry of noble metal oxides. I. Syntheses and properties of ABO$_2$ delafossite compounds. \textit{Inorg. Chem. }{\bf 10}, 713 (1971).

\bibitem{Prewitt} C. T. Prewitt, R. D. Shannon, and D. B. Rogers, Chemistry of noble metal oxides. II. Crystal structures of platinum cobalt dioxide, palladium cobalt dioxide, coppper iron dioxide, and silver iron dioxide. \textit{Inorg. Chem.} {\bf 10}, 719 (1971).

\bibitem{Rogers} D. B. Rogers, R. D. Shannon, C. T. Prewitt, and J. L. Gillson, Chemistry of noble metal oxides. III. Electrical transport properties and crystal chemistry of ABO$_2$ compounds with the delafossite structure. \textit{Inorg. Chem.} {\bf 10}, 723 (1971).

\bibitem{Marquardt} M. A. Marquardt, N. A. Ashmore, and D. P. Cann,Crystal chemistry and electrical properties of the delafossite structure.\textit{ Thin Solid Films} {\bf 496}, 146 (2006).

\bibitem{Singh} D. J. Singh,Electronic and thermoelectric properties of CuCoO$_2$: Density functional calculations.\textit{ Phys. Rev. B} {\bf 76}, 085110 (2007).

\bibitem{Wang} F. Wang and A. Vishwanath, Spin Phonon Induced Collinear Order and Magnetization Plateaus in Triangular and Kagome Antiferromagnets: Applications to CuFeO$_2$. \textit{Phys. Rev. Lett.} {\bf 100}, 077201 (2008).

\bibitem{Yanagi} H. Yanagi, T. Hase, S. Ibuki, K. Ueda, and H. Hosono, Bipolarity in electrical conduction of transparent oxide semiconductor CuInO$_2$ with delafossite structure. \textit{Appl. Phys. Lett.} {\bf 78}, 1583 (2001), and references therein.

\bibitem{Takatsu 2009} H. Takatsu, H. Yoshizawa, S. Yonezawa and Y. Maeno,Critical behavior of the metallic triangular-lattice Heisenberg antiferromagnet PdCrO$_2$. \textit{Phys. Rev. B} {\bf 79}, 104424 (2009).

\bibitem{Takatsu 2010} H. Takatsu, S. Yonezawa, S. Fujimoto, and Y. Maeno, Unconventional Anomalous Hall Effect in the Metallic Triangular-Lattice Magnet PdCrO$_2$. \textit{Phys. Rev. Lett}. {\bf 105}, 137201 (2010).

\bibitem{Ok} J. M. Ok, Y. J. Jo, K. Kim, T. Shishidou, E. S. Choi, H. J. Noh, T.  Oguchi, B. I. Min, and J. S. Kim, Quantum Oscillations of the Metallic Triangular-Lattice Antiferromagnet PdCrO$_2$. \textit{Phys. Rev. Lett.} {\bf 111}, 176405 (2014).

\bibitem{Takatsu 2014} H. Takatsu, G. Nenert, H. Kadowaki, H. Yoshizawa, M. Enderle, S. Yonezawa, Y. Maeno, J. Kim, N. Tsuji, M. Takata, Y. Zhao, M. Green and C. Broholm, Magnetic structure of the conductive triangular-lattice antiferromagnet PdCrO$_2$. \textit{Phys. Rev. B} {\bf 89}, 104408 (2014).

\bibitem{Hicks 2015} arXiv:1504.08104v1
Quantum Oscillations and Magnetic Reconstruction in the Delafossite PdCrO$_2$.
Clifford W. Hicks, Alexandra S. Gibbs, Lishan Zhao, Pallavi Kushwaha, Horst Borrmann, Andrew P.
Mackenzie, Hiroshi Takatsu, Shingo Yonezawa, Yoshiteru Maeno, and Edward A. Yelland.

\bibitem{Coldea} A. I. Coldea, L. Seabra, A. McCollam, A. Carrington, L. Malone, A. F. Bangura, D. Vignolles, P. G. van Rhee, R. D. McDonald, T. Sorgel, M. Jansen, N. Shannon, and R. Coldea,Cascade of field-induced magnetic transitions in a frustrated antiferromagnetic metal. \textit{Phys. Rev. B} {\bf 90}, 020401 (2014).

\bibitem{Eyert} V. Eyert, R. Fr\'{e}sard, and A. Maignan,On the Metallic Conductivity of the Delafossites PdCoO$_2$ and PtCoO$_2$. \textit{Chem. Mater.} {\bf 20}, 2370 (2008).

\bibitem{Seshadri}R. Seshadri, C. Felser, K. Thieme, and W. Tremel, Metal-Metal Bonding and Metallic Behavior in Some ABO$_2$ Delafossites. \textit{Chem. Mater.} {\bf 10}, 2189 (1998).

\bibitem{Kim} K. Kim, H. C. Choi, and B. I. Min, Fermi surface and surface electronic structure of delafossite PdCoO$_2$. \textit{Phys. Rev. B }{\bf 80}, 035116 (2009).

{\color{black}\bibitem{Noh_orb} H-J. Noh, J. Jeong, J. Jeong, H. Sung, K. J. Park, J.-Y. Kim, H.-D. Kim, S. B. Kim, K. Kim, and B. I. Min, Orbital character of the conduction band of delafossite PdCoO$_2$ studied by polarization-dependent soft x-ray absorption spectroscopy. \textit{Phys. Rev. B} {\bf 80}, 073104 (2009).}

\bibitem{Hicks} C. W. Hicks, A. S. Gibbs, A. P. Mackenzie, H. Takatsu, Y. Maeno and E. A. Yelland, Quantum Oscillations and High Carrier Mobility in the Delafossite PdCoO$_2$. \textit{Phys. Rev. Lett.} {\bf 109}, 116401 (2012).

\bibitem{Takatsu 2013}  H. Takatsu, J. J.  Ishikawa, S. Yonezawa, H. Yoshino, T. Shishidou, T. Oguchi, K. Murata, Y. Maeno, Extremely Large Magnetoresistance in the Nonmagnetic Metal PdCoO$_2$. \textit{Phys. Rev. Lett.} {\bf 111}, 056601 (2013).

\bibitem{Kikugawa 2014}  arXiv:1412.5168 
Realization of the axial anomaly in a quasi-two-dimensional metal
N. Kikugawa, P. Goswami, A. Kiswandhi, E. S. Choi, D. Graf, R. E. Baumbach, J. S. Brooks, K. Sugii, Y. Iida, M. Nishio, S. Uji, T. Terashima, P. M. C. Rourke, N. E. Hussey, H. Takatsu, S. Yonezawa, Y. Maeno, L. Balicas

\bibitem{Goswami 2015} arXiv:1503.02069 
Axial anomaly and longitudinal magnetoresistance of a generic three dimensional metal
Pallab Goswami, J. H. Pixley, S. Das Sarma


\bibitem{uspatent} R. D. Shannon, Electrically conductive platinum cobalt oxides. \textit{US patent} 3,514,414 (1970).

\bibitem{Tanaka} M. Tanaka, M. Hasegawa, H. Takei, Crystal growth of PdCoO$_2$, PtCoO$_2$ and their solid-solution with delafossite structure.\textit{ J. Cryst. Growth} {\bf 173}, 440 (1997).

\bibitem{Takatsu 2007} H. Takatsu, S. Yonezawa, S. Mouri, S. Nakatsuji, K. Tanaka, and Y. Maeno, Roles of High-Frequency Optical Phonons in the Physical Properties of the Conductive Delafossite PdCoO$_2$. \textit{J. Phys. Soc. Jpn.}, {\bf 76}, 104701 (2007).

\bibitem{Pallavi2015} Pallavi Kushwaha, C. Geibel, M. P. Schmidt, A. P. Mackenzie et al., to be published.

\bibitem{Moll}  P. J. W. Moll, R. Puzniak, F. Balakirev, K.  Rogacki, J. Karpinski, N.D. Zhigadlo and  B.  Batlogg, High magnetic-field scales and critical currents in SmFeAs(O, F) crystals. \textit{Nature Materials} {\bf 9}, 628 (2010).

{\color{black}
\bibitem{Pearsall} T. Pearsall and C. Lee, Electronic transport in ReO$_3$: dc conductivity and Hall effect. \textit{Phys. Rev. B} {\bf 10}, 2190 (1974).

\bibitem{Ryden} W. D. Ryden, A. W. Lawson, and C. C. Sartain, Temperature dependence of the resistivity of RuO$_2$ and IrO$_2$. \textit{Phys. Lett.}, {\bf 26A}, 209 (1968).
}

\bibitem{Hurd} C. M. Hurd, The Hall effect in metals and alloys, Springer US; Boston, MA (1972).

\bibitem{Riley} J. M. Riley, F. Mazzola, M. Dendzik,	M. Michiardi, T. Takayama, L. Bawden, C. Granerd, M. Leandersson, T. Balasubramanian, M. Hoesch, T. K. Kim, H. Takagi, W. Meevasana, Ph. Hofmann, M. S. Bahramy,  J. W. Wells, and P. D. C. King, Direct observation of spin-polarized bulk bands in an inversion-symmetric semiconductor. \textit{Nature Phys.} {\bf 10}, 835 (2014).

\bibitem{Noh} H. -J. Noh, J. Jeong, J. Jeong, E. -J. Cho, S. B. Kim, Kyoo Kim, B. I. Min, and H. -D. Kim,
Anisotropic Electric Conductivity of Delafossite PdCoO$_2$ Studied by Angle-Resolved Photoemission Spectroscopy. \textit{Phys. Rev. Lett.} {\bf 102}, 256404 (2009).

\bibitem{Ong and Singh} K. P. Ong, D. J. Singh, and P. Wu, Unusual Transport and Strongly Anisotropic Thermopower in PtCoO$_2$ and PdCoO$_2$. \textit{Phys. Rev. Lett.} {\bf 104}, 176601 (2001).

{\color{black}\bibitem{Opos} C. T. Prewitt, R. D. Shannon, and D. B. Rogers, Chemistry of Noble Metal Oxides. II. Crystal Structures of PtCoO$_2$, PdCoO$_2$, CuFeO$_2$, and AgFeO$_2$. \textit{Inorganic Chemistry} {\bf 10}, 719 (1971).}

\bibitem{dHvA} F.~Arnold, E.~Hassinger et al., to be published.

\bibitem{Sobota} J. A. Sobota, K. Kim, H. Takatsu, M. Hashimoto, S.-K. Mo, Z. Hussain, T. Oguchi, T. Shishidou, Y. Maeno, B. I. Min, and Z.-X. Shen, Electronic structure of the metallic antiferromagnet PdCrO$_2$ measured by angle-resolved photoemission spectroscopy. \textit{Phys. Rev. B} {\bf 88}, 125109 (2013). 

\bibitem{Noh2} H. -J. Noh, J. Jeong, B. Chang, D. Jeong, H. S. Moon, E. -J. Cho, J. M. Ok, J. S. Kim, K. Kim, B. I. Min, H.-K. Lee, J.-Y. Kim, B.-G. Park, H. -D. Kim, and S. Lee, Direct Observation of Localized Spin Antiferromagnetic Transition in PdCrO$_2$ by Angle-Resolved Photoemission Spectroscopy. \textit{Sci. Report} {\bf 4}, 3680 (2014).

\bibitem{dlT} A. de la Torre, E. C. Hunter, A. Subedi, S. McKeown Walker, A. Tamai, T. K. Kim, M. Hoesch, R. S. Perry, A. Georges, and F. Baumberger, Coherent Quasiparticles with a Small Fermi Surface in Lightly Doped Sr$_3$Ir$_2$O$_7$. \textit{Phys. Rev. Lett.} {\bf 113}, 256402 (2014).

\bibitem{Nie} Y. F. Nie, P. D. C. King, C. H. Kim, M. Uchida, H. I. Wei, B.D. Faeth, J.P. Ruf, J.P.C. Ruff, L. Xie, X. Pan, C.J. Fennie, D.G. Schlom, and K. M. Shen, Interplay of Spin-Orbit Interactions, Dimensionality, and Octahedral Rotations in Semimetallic SrIrO$_3$. \textit{Phys. Rev. Lett.}  {\bf 114}, 016401 (2015).

\bibitem{TanakaPB} Masayuki Tanakaa,Masashi Hasegawaa, Thoru Higuchib, Takeyo Tsukamotob, Yasuhisa Tezukaa, Shik Shina, Humihiko Takeic, Origin of the metallic conductivity in PdCoO$_2$ with delafossite structure. \textit{Physica B} {\bf 245}, 157 (1998).

\bibitem{Ketterson} J. B. Ketterson,  and L. R. Windmiller, de Haas-van Alphen Effect in Platinum. \textit{Phys. Rev. B}, {\bf 2}, 4813 (1970).

\bibitem{Damascelli} A. Damascelli, D. H. Lu, K. M. Shen, N. P. Armitage, F. Ronning, D. L. Feng, C. Kim, Z.-X. Shen, T. Kimura, Y. Tokura, Z. Q. Mao, and Y. Maeno,Fermi Surface, Surface States, and Surface Reconstruction in Sr$_2$RuO$_4$. \textit{Phys. Rev. Lett.} {\bf 85}, 5194 (2000).

\bibitem{Baumberger} F. Baumberger, N. J. C. Ingle, W. Meevasana, K. M. Shen, D. H. Lu, R. S. Perry, A. P. Mackenzie, Z. Hussain, D. J. Singh, and Z.-X. Shen, Fermi Surface and Quasiparticle Excitations of Sr$_2$RhO$_4$. \textit{Phys. Rev. Lett.} {\bf 96}, 246402 (2006).

\bibitem{BJKim} B. J. Kim, Hosub Jin, S. J. Moon, J.-Y. Kim, B.-G. 
Park, C. S. Leem, Jaejun Yu, T. W. Noh, C. Kim, S.-J. Oh, J.-H. Park, V. Durairaj, G. Cao, and E. Rotenberg, Novel ${J}_{eff}$=1/2 Mott State Induced by Relativistic Spin-Orbit Coupling in Sr$_2$IrO$_4$. \textit{Phys. Rev. Lett.}, {\bf 101}, 076402 (2008).

\bibitem{King} P. D. C. King, T. Takayama, A. Tamai, E. Rozbicki, S. McKeown Walker, M. Shi, L.~Patthey, R. G. Moore, D. Lu, K. M. Shen, H. Takagi, and F. Baumberger, Spectroscopic indications of polaronic behavior of the strong spin-orbit insulator Sr$_3$Ir$_2$O$_7$. \textit{Phys. Rev. B}, {\bf 87}, 241106(R) (2013).

\bibitem{fplo} K. Koepernik, and H. Eschrig, Full-potential nonorthogonal local-orbital minimum-basis band-structure scheme. \textit{Phys. Rev. B}, {\bf 59}, 1743 (1999); I. Opahle, K. Koepernik, and H. Eschrig, Full-potential band-structure calculation of iron pyrite. \textit{Phys. Rev. B}, {\bf 60}, 14035 (1999); http://www.fplo.de

\bibitem{PBE} J.P. Perdew, K. Burke and M. Ernzerhof, Generalized Gradient Approximation Made Simple. \textit{Phys. Rev. Lett.}, {\bf 77}, 3865 (1996).

{\color{black}\bibitem{KSD} H. Eschrig, M. Richter and I. Opahle, Relativistic Solid State Calculations, in: "Relativistic Electronic Structure Theory, (Part II, Applications)" (ed. P. Schwerdtfeger), Theoretical and Computational Chemistry, vol.13, Elsevier, 723, (2004)}

\bibitem{LDU} H. Eschrig, K. Koepernik, and I. Chaplygin, Density functional application to strongly correlated electron systems. \textit{J. Solid State Chemistry}, {\bf176}, 482 (2003)

%\bibitem{Andreev} A. V. Andreev, S. A. Kivelson and B. Spivak,Hydrodynamic Description of Transport in Strongly Correlated Electron Systems. \textit{Phys. Rev. Lett.} {\bf 106}, 256804 (2011).

{\color{black}\bibitem{Ong} N. P. Ong, Geometric interpretation of the weak-field Hall conductivity in two-dimensional metals with arbitrary Fermi surface. \textit{Phys. Rev. B} {\bf 43}, 193 (1991).}

\end{thebibliography}

\pagebreak

 \begin{widetext}
\onecolumngrid
\begin{center}
\textbf{\large Supplementary Material: Nearly-free electrons in a $5d$ delafossite oxide metal}

\title{Nearly-free electrons in a $5d$ delafossite oxide metal}
\end{center}
\vskip\baselineskip
 \end{widetext}

\setcounter{equation}{0}
\setcounter{figure}{0}
\setcounter{table}{0}
\setcounter{page}{1}
\makeatletter
\renewcommand{\theequation}{S\arabic{equation}}
\renewcommand{\thefigure}{S\arabic{figure}}
\renewcommand{\bibnumfmt}[1]{[S#1]}
\renewcommand{\citenumfont}[1]{S#1}
\makeatother

\section*{Hall Effect Measurement}
\label{sec:dejong}
\begin{figure}[!b]
	\begin{center}
	\includegraphics[width=10 cm]{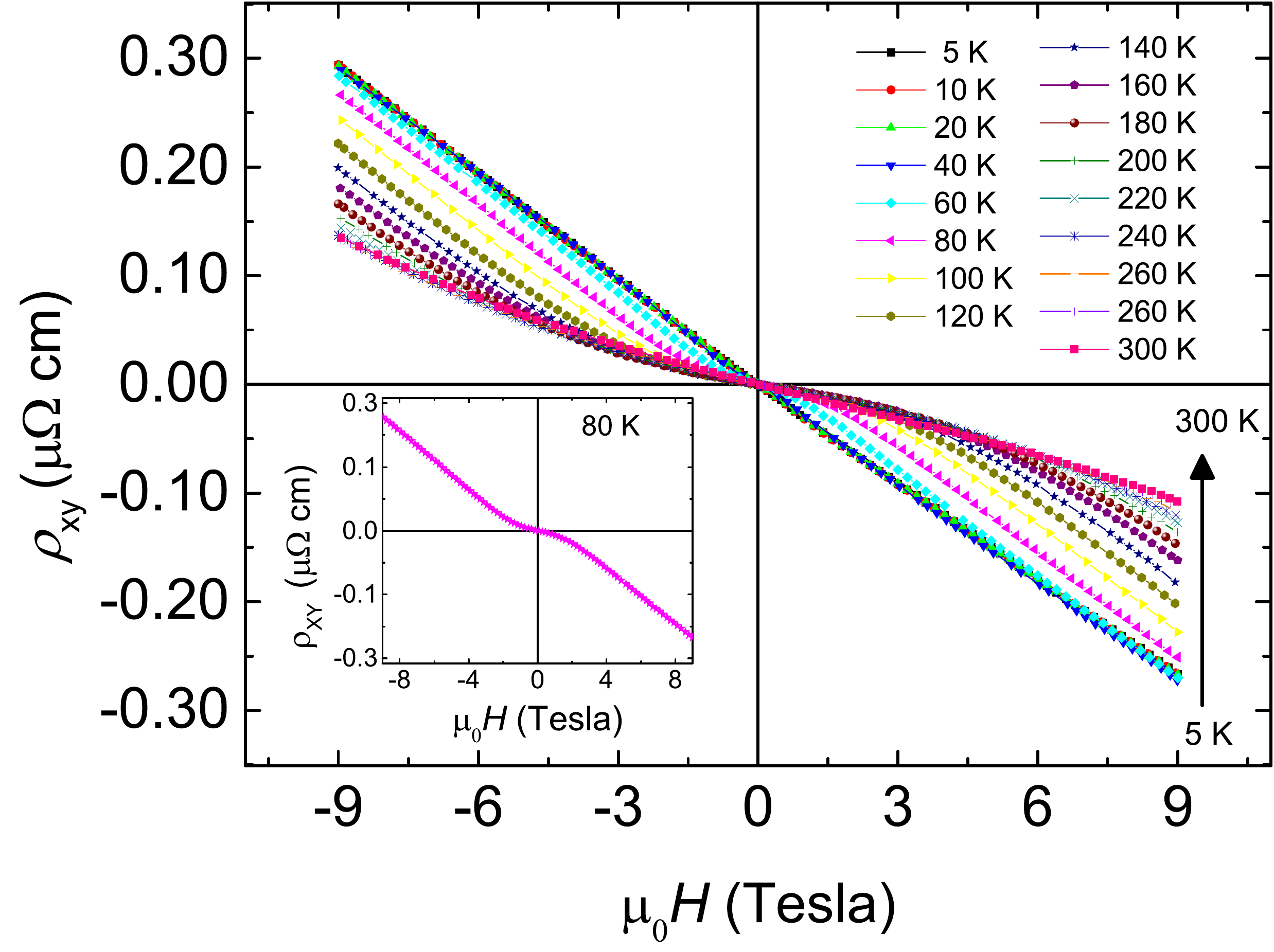}
	\end{center}
	\caption{\label{sf1} Field dependence of Hall resistivity ($\rho_{xy}$) for PtCoO$_2$ for different labeled temperatures for the $2.6~\mu$m thick sample. The inset highlights the two regions seen as a function of applied magnetic field, using data taken at 80~K as an example.}
\end{figure}

The Hall coefficient shown in the inset of Fig. 2 of the main text were obtained from field sweeps performed at a number of temperatures are shown in Fig.~\ref{sf1}.  We assume that the Hall effect in non-magnetic PtCoO$_2$ or PdCoO$_2$ has no anomalous component~[9], and that the signal is dominated by the orbital Hall effect. There is a clear, temperature-dependent separation into low- and high-field regimes with different gradients of the Hall resistivity $\rho_{xy}$.  Below 20~K, the data are dominated by the high-field regime in which the Fermi surface volume alone is expected to determine the value of the Hall coefficient~[29]. The value in that region is approximately 2.9$\times 10^{-10}$~m$^3$/C, close to the value that we measure for PdCoO$_2$ under the same conditions (data not shown), and within $10~\%$ of the expectation for a single band containing one electron per Pd. Our measured mean-free path (see main text) enables us to make a quantitative estimate of $l/$r$_c$ = $\omega$$_c\tau$ at each temperature and field (here the cyclotron radius $r_c$ = $\hbar$$k_{\mathrm{F}}/eB$, the cyclotron frequency $\omega_c$= $eB/m^{*}$ = $eBv_{\mathrm{F}}/$$\hbar$$k_{\mathrm{F}}$ and $\textit{l}$ = $v_{\mathrm{F}}$$\tau$ where $\tau$ is the relaxation time). In fields of 9 T, we find $l/$r$_c$ $\geq$~5 at 15~K, which drops to $l/$r$_c$ = 0.1 at 300~K. This indicates, therefore, that the experiment goes well beyond the weak field regime by 9 T at low temperatures, but that weak-field physics applies at all fields at high temperatures.  Accordingly, we observe a cross-over from the low-field to the high-field regime ($l/$r$_c$~$\approx$~1) at low temperatures, while at elevated temperatures fields in excess of 9 T are required to observe the high-field limit. We attribute this crossover between weak- and strong-field physics as the origin of two different gradients at intermediate temperatures (see for example the inset to Fig.~\ref{sf1}). The Hall coefficient data shown in inset to Fig. 2 of the main paper were obtained by fitting the gradient of $\rho_{xy}$ between 7 and 9~T. This ensures that below approximately 50~K, R$_H$ is obtained from data in the high filed limit that is more directly related to Fermi surface volume that data from weak or intermediate fields. We attribute the temperature dependence seen above 50~K to the crossover to those less reliable regimes in which details of \textit{k}-dependent scattering strongly effect R$_H$~[49].

\section*{Calculated density of states}
\label{sec:dejong}
\begin{figure}
	\begin{center}
	\includegraphics[width=12 cm]{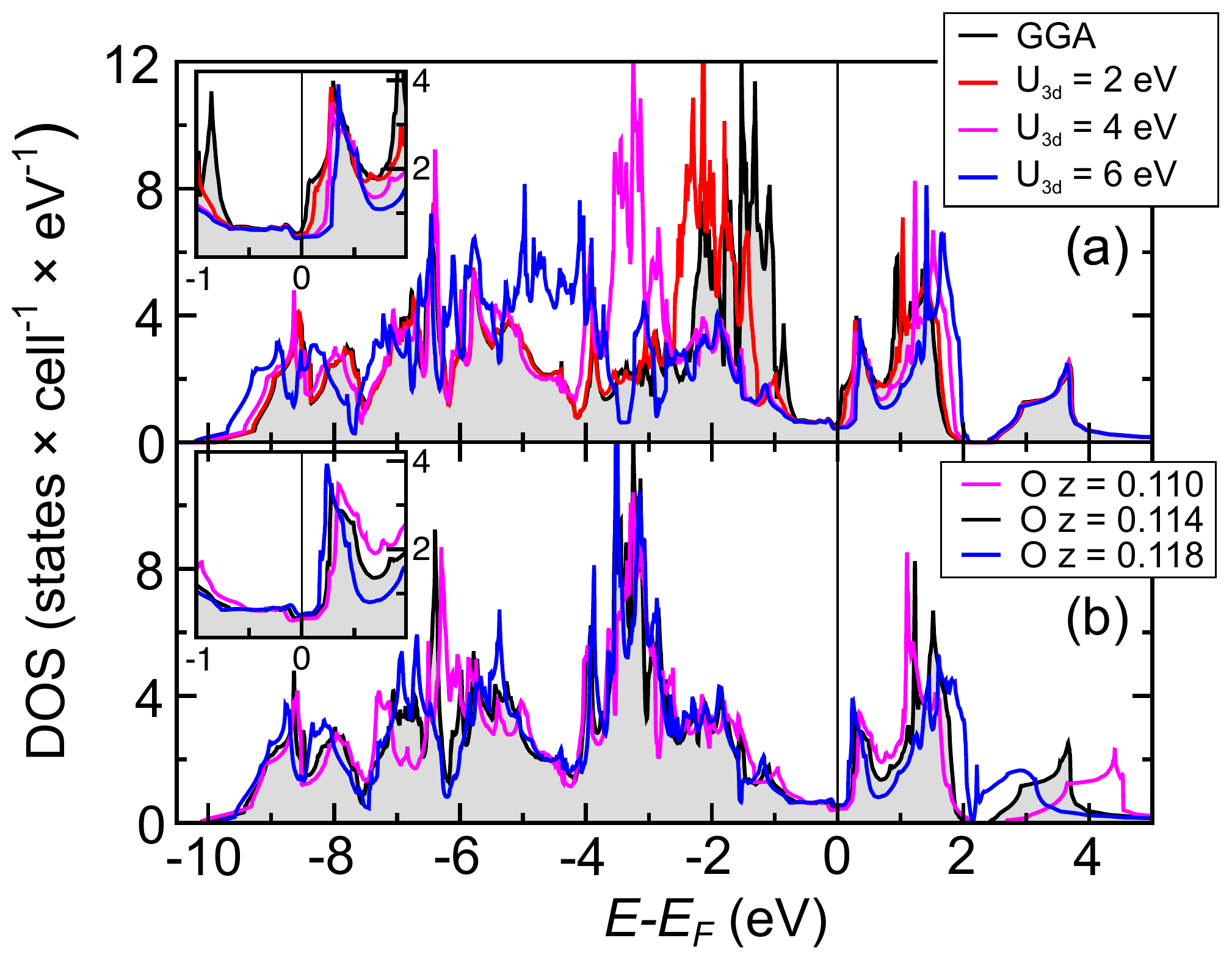}
	\end{center}
	\caption{\label{sf2}  Calculated density of states (DOS). Dependence of the DOS on (a) the Coulomb repulsion $U_{3d}$ in the Co 3d shell for $z=0.114$ and (b) the O-position (0,0,$z$) for $U_{3d}=4$~eV. The insets show magnified views near the Fermi energy, $E_F$, and all calculations include spin-orbit coupling.}
\end{figure}

Supplemental Fig.~\ref{sf2} shows the dependence of the calculated density of states on the Coulomb repulsion $U_{3d}$ in the Co 3d shell and the oxygen position  within the crystal structure. Explicitly taking into account the strong Coulomb repulsion $U_{3d}$ suppresses unphysical hybridization of the Co-$3d$ states with the Pt-$5d$ states crossing the Fermi level but otherwise has a minor influence on the DOS in the vicinity of $E_F$. In contrast, it has a significant influence at higher binding energy where the valence bands have significant Co orbital character as shown in Fig.~4(b) of the main text. Varying the O-position (0,0,$z$) induces a shift of the energy of a hole band whose band top is slightly below $E_F$, giving rise to a shift of a step in the density of states visible in the inset of Fig.~\ref{sf2}(b). In the absence of spin-orbit interactions, this would cause the hole band to move above $E_F$ for Pt-O distances at and above their calculated relaxed value ($z > 0.112$; the experimental value is z=0.114) causing a second Fermi surface to appear. However, this is pushed below $E_F$ for any realistic value of $z$ when including spin-orbit coupling in the calculation, as observed in Fig.~\ref{sf2}(b).

\end{document}